\documentclass[11pt, a4paper]{article}
\usepackage{amsmath,amssymb}
\usepackage{float}
\usepackage{graphicx}
\usepackage{subfigure}
\usepackage{amssymb}
\usepackage{color}
\newtheorem{theorem}{Theorem}

\begin{document}

\title{Characteristic integrals in 3D and linear degeneracy}
\author{E. V. Ferapontov and J. Moss
}
   \date{}
\vspace{-20mm}
   \maketitle
\vspace{-7mm}
\begin{center}
Department of Mathematical Sciences \\ Loughborough
University \\
Loughborough, Leicestershire LE11 3TU, UK \\[1ex] 
e-mails: \\
\texttt{E.V.Ferapontov@lboro.ac.uk}\\
\texttt{J.Moss@lboro.ac.uk}\\
\end{center}

\medskip

\begin{abstract}
\noindent Conservation laws vanishing along characteristic directions of a given system of PDEs are known as characteristic conservation laws, or characteristic integrals. In 2D, they play an important role in the theory of Darboux-integrable equations. In this paper we discuss characteristic integrals in  3D and demonstrate that, for a class of second-order  linearly degenerate dispersionless  integrable PDEs, the corresponding characteristic integrals are parametrised by points on the Veronese variety.


\medskip

\end{abstract}
 
\newpage

\section{Introduction} 

Let $\Sigma$ be a  partial differential equation (PDE) in $n$ independent variables $x_1, \dots, x_n$. A conservation law  is an $(n-1)$-form $\Omega$ which is closed on the solutions of $\Sigma$: $d\Omega=0 {\rm ~ mod ~} \Sigma$.  Since any $(n-1)$-form in $n$ variables possesses a unique annihilating direction, there exists a  vector field  $F$ such  that $\Omega (F)=0$. We say that $\Omega$ is a characteristic integral (conservation law)  if $F$ is a characteristic direction of $\Sigma$. \footnote{The set of characteristic directions is projectively dual to the more conventional variety of  characteristic covectors determined by the principal symbol of the equation. } If a conservation law is represented in  conventional form, 
$$
(F_1)_{x_1}+\dots + (F_n)_{x_n}=0 {\rm ~ mod ~} \Sigma,
$$
the corresponding vector field is $F=(F_1, \dots, F_n)$. The characteristic condition becomes particularly simple for scalar second-order PDEs, in which case $F$ can be interpreted as a null vector of the conformal structure defined by the  principal symbol of the equation. Let us begin with illustrating examples.

\medskip

\noindent {\bf Example 1.}  Consider the $2+1$ dimensional wave equation,
\begin{equation}
u_{tt} = u_{xx}+u_{yy}.
\label{waveeqn}
\end{equation}
It possesses four first-order conservation laws, 
$$
(u_x)_x+(u_y)_y-(u_t)_t=0,
$$
$$
(u_x^2+u_t^2-u_y^2)_x+(2u_xu_y)_y-(2u_xu_t)_t=0,
$$
$$
(2u_yu_x)_x-(u_x^2-u_y^2-u_t^2)_y-(2u_yu_t)_t=0,
$$
$$
(2u_tu_x)_x+(2u_tu_y)_y-(u_x^2+u_y^2+u_t^2)_t=0.
$$
Let us denote them 
$$
(f_i)_{x}+(g_i)_{y}+(h_i)_{t}=0, 
$$ 
$ i=1, ..., 4$.  Taking their linear combination with constant coefficients $J_1, \dots, J_4$, and adding trivial conservation laws,
we obtain the expression $(F_1)_{x}+(F_2)_{y} + (F_3)_{t}=0$ where
$$
F_1=J_i f_i - J_5 u_{y} + J_6 u_{t} + J_8, ~~~ F_2= J_i g_i + J_5 u_{x} - J_7 u_{t} + J_9, ~~~
F_3=J_i h_i - J_6 u_{x} + J_7 u_{y} + J_{10},
$$
(summation over $i=1, \dots, 4$ is assumed). Here the constants $J_5, J_6, J_7$ correspond to trivial conservation laws of the form $(u_{x})_{y}-(u_{y})_{x}=0$, etc., and $J_8, J_9, J_{10}$ are three extra arbitrary constants. 
Although the constants $J_5 - J_{10}$ correspond to trivial conservation laws, they effect non-trivially the characteristic condition, $Fg^{-1}F^t=0$,
where $g$ is the $3\times 3$ symmetric matrix of the corresponding principal symbol,
$$
g=
\left(
\begin{array}{ccc}
-1 & 0 & 0 \\
0  & -1 & 0\\
 0 & 0 & 1 \end{array}\right),
$$
(in this particular example $g$ coincides with $g^{-1}$). The characteristic condition imposes a system of quadratic constraints for  $J_1, \dots, J_{10}$, which specify a Veronese threefold $V^3\subset P^9$ with parametric equations
$$
J_1=\frac{\sqrt{2}}{2}(\gamma \alpha + \gamma \beta + \delta \alpha - \delta \beta), \hspace{3mm} J_2=-\alpha \beta, \hspace{3mm} J_3=\frac{\alpha^2-\beta^2}{2}, \hspace{3mm} J_4=-\frac{\alpha^2+\beta^2}{2},
$$
$$
J_5=\frac{\sqrt{2}}{2}(\gamma \alpha - \gamma \beta - \delta \alpha - \delta \beta), \hspace{3mm} J_6=-\frac{\sqrt{2}}{2}(\gamma \alpha +\gamma \beta -\delta \alpha +\delta \beta),
$$
$$
J_7=\frac{\sqrt{2}}{2}(\gamma \alpha - \gamma \beta +\delta \alpha +\delta \beta), \hspace{3mm} J_8=\frac{\delta^2-\gamma^2}{2}, \hspace{3mm} J_9=\delta \gamma, \hspace{3mm} J_{10}=\frac{\gamma^2+\delta^2}{2}.
$$
We use $\alpha, \beta, \gamma,  \delta$ as homogeneous coordinates in $P^3$, and $J_1, \dots, J_{10}$ as homogeneous coordinates in $P^9$. Recall that the Veronese threefold $V^3$  is the image of the projective imbedding of  $P^3$ into $P^9$ defined by a complete system of quadrics. Thus, we have a whole $V^3$-worth of characteristic integrals. It turns out that this example is not isolated, and  similar phenomena take place for  other classes of 3D linearly degenerate dispersionless integrable PDEs.

\medskip

\noindent{\bf Example 2.} Let us consider the equation
\begin{equation}
\mu u_{t}u_{xy}+\nu u_{y} u_{xt}+\eta u_{x}u_{yt}=0,
\label{Ver}
\end{equation}
$\mu+\nu+\eta=0$, which  appeared  in the context of Veronese webs in 3D  \cite{Zakharevich}, as well as in the theory of Einstein-Weyl geometries of  hyper-CR type \cite{Dun1}. This equation possesses four first-order conservation laws,
$$
\eta (u_{y}u_{t})_{x}+\nu(u_{x}u_{t})_{y}+\mu(u_{x}u_{y})_{t}=0,
$$
$$
\nu \left(\frac{u_{y}}{u_{t}}\right)_{x}+\eta \left(\frac{u_{x}}{u_{t}}\right)_{y}=0,
$$
$$
\mu \left(\frac{u_{t}}{u_{y}}\right)_{x}+\eta \left(\frac{u_{x}}{u_{y}}\right)_{t}=0,
$$
$$
\mu \left(\frac{u_{t}}{u_{x}}\right)_{y}+\nu \left(\frac{u_{y}}{u_{x}}\right)_{t}=0.
$$
Let us denote them 
$$
(f_i)_{x}+(g_i)_{y}+(h_i)_{t}=0, 
$$ 
$ i=1, ..., 4$.  Taking their linear combination with  coefficients $J_1, \dots, J_4$, and adding trivial conservation laws,
we obtain the expression $(F_1)_{x}+(F_2)_{y} + (F_3)_{t}=0$ where
$$
F_1=J_i f_i - J_5 u_{y} + J_6 u_{t} + J_8, ~~~ F_2= J_i g_i + J_5 u_{x} - J_7 u_{t} + J_9, ~~~
F_3=J_i h_i - J_6 u_{x} + J_7 u_{y} + J_{10},
$$
As in Example 1,  the constants $J_5 - J_{10}$ correspond to trivial conservation laws. The characteristic condition takes the form $Fg^{-1}F^t=0$
where $g$ is the $3\times 3$ symmetric matrix of the corresponding principal symbol:
$$
g=
\left(
\begin{array}{ccc}
0& \mu u_{t} &\nu u_{y}\\
 \mu u_{t} &0& \eta u_{x}\\
 \nu u_{y}& \eta u_{x}&0\end{array}\right).
$$
The characteristic condition imposes a system of quadratic constraints for $J_1, \dots,  J_{10}$, which specify a Veronese threefold $V^3\subset P^9$ with parametric equations
$$
J_1=\alpha^2, \hspace{3mm} J_2=\frac{1}{4 \nu \eta}\beta^2, \hspace{3mm} J_3=\frac{1}{4 \eta \mu}\delta^2, \hspace{3mm} J_4=\frac{1}{4 \nu \mu}\gamma^2,
$$
$$
J_5=\alpha \beta,  \hspace{3mm}  J_6=\alpha \delta, \hspace{3mm} J_7=\alpha \gamma, \hspace{3mm} J_8=-\frac{1}{2 \eta} \beta \delta, \hspace{3mm} J_9=-\frac{1}{2 \nu}\beta \gamma, \hspace{3mm} J_{10}=-\frac{1}{2\mu}\delta\gamma.
$$
Further examples of this type are given in the Appendix.
\medskip

The structure of the paper is as follows: 

\noindent In Sect. 2 we briefly review the case of  2D hydrodynamic type systems, where the existence of characteristic integrals is known to imply linear degeneracy.


\noindent In Sect. 3 we consider characteristic integrals of second-order  quasilinear PDEs in 3D. We prove that the requirement of the existence of `sufficiently many' integrals of this type implies linear degeneracy. For all linearly degenerate integrable cases,  we obtain $V^3$-worth of characteristic integrals.


\medskip

In two dimensions, characteristic integrals arise in the context of Darboux integrability, see e.g. \cite{Sokolov, Zhiber} and references therein. We thus hope that the results of this paper may be useful for the theory of multi-dimensional Darboux-integrable equations (yet to be constructed).  We refer to  \cite{Parsons, Hartley, Kakie75, Athorne95, Athorne12} for some  steps in this direction.

\section{Characteristic integrals in 2D and linear degeneracy}

For definiteness we restrict the discussion to systems of  hydrodynamic type,
\begin{equation}
u^i_t=v^i_j({\bf u})u^i_x,
\label{ReInv}
\end{equation}
where ${\bf u}=(u^1, \dots, u^n)$ denotes dependent variables, and $v=v^i_j$ is an $n\times n $ matrix. Let $\lambda^i$ be the eigenvalues (characteristic speeds) of $v$, and let $\xi^i$ be the corresponding eigenvectors, so that 
$
v\xi^i=\lambda^i \xi^i.
$
Characteristic directions  are defined as $dx+\lambda^i dt=0$, and the  characteristic integral in  $i$-th direction is a 1-form $h({\bf u})(dx+\lambda^i dt)$ which is closed on solutions of (\ref{ReInv}). We will assume that the density $h$  depends on  ${\bf u}$ only, although, in principle, nontrivial dependence on higher-order $x$-derivatives of ${\bf u}$ may also be allowed. Recall that the $i$-th characteristic direction is called linearly degenerate if the Lie derivative of $\lambda^i$ in the direction of the corresponding eigenvector $\xi^i$ vanishes, $L_{\xi^i}\lambda^i=0$.
The following result is well-known:

\medskip

\noindent {\bf Proposition.} {\it If there exists a characteristic integral in the $i$-th direction, then the corresponding characteristic speed $\lambda^i$ must be linearly degenerate.}

\medskip

\centerline{\bf Proof:}

\medskip

The closedness of $h({\bf u})(dx+\lambda^i dt)$ is equivalent to 
$
h_t=(\lambda^ih)_x.
$
This implies 
$$
(\nabla h) v=h\nabla \lambda^i+\lambda^i\nabla h,
$$
where $\nabla=(\partial_{u^1}, \dots, \partial_{u^n})$ denotes the gradient.  Evaluating both sides of this identity (which are 1-forms)  on the vector 
$\xi^i$, and using $v\xi^i=\lambda^i \xi^i$, one can see that the left hand side cancels with the second term on the right hand side,  leading to $\xi^i \nabla \lambda^i = L_{\xi^i}\lambda^i=0$. This finishes the proof.

The requirement of the existence of characteristic integrals for all characteristic directions implies that all characteristic speeds must be linearly degenerate. Such systems are known as (totally) linearly degenerate, they  have been thoroughly investigated in the literature, see e.g. \cite{R1, R2, Liu, Serre}. For linearly degenerate systems the gradient catastrophe, which is typical for genuinely nonlinear systems, does not occur, and one has global existence results for an open set of  initial data.

There exist systems which possess infinitely many characteristic integrals. 

\noindent{\bf Example 3.} The 2-component linearly degenerate  system,
$$
v_t=wv_x, \hspace{5mm} w_t=vw_x,
$$ 
possesses functionally many  characteristic integrals in both characteristic directions:
$$
\frac{\phi(v)}{w-v}(dx+wdt), ~~~~~~ \frac{\psi(w)}{v-w}(dx+vdt),
$$
here $\phi$ and $\psi$ are arbitrary functions of $w$ and $v$ respectively.

\section{Characteristic integrals of second-order   PDEs in 3D}

In this section we consider quasilinear wave-type equations of the form
 \begin{equation}
f_{11} u_{xx} + f_{22} u_{yy} + f_{33} u_{tt} + 2 f_{12} u_{xy} + 2f_{13} u_{xt} + 2f_{23} u_{yt} =0, 
 \label{u}
 \end{equation}
where $u(x, y, t)$ is a function of three independent variables, and the coefficients 
$f_{ij}$ depend on  the first-order derivatives  $u_{x}, u_{y}, u_{t}$ only. Equations of this form generalise Examples 1, 2 from the introduction.
It was shown in \cite{B} that any integrable  equation of the form (\ref{u}) possesses exactly four conservation laws  
$$
(f_i)_{x}+(g_i)_{y}+(h_i)_{t}=0,
$$
$i=1, \dots, 4$, where $f_i, g_i, h_i$ are functions of $u_{x}, u_{y}, u_{t}$ only. Taking their linear combination with constant coefficients $J_1, \dots, J_4$, and adding trivial conservation laws,
we obtain the expression $(F_1)_{x}+(F_2)_{y} + (F_3)_{t}=0$ where
$$
F_1=J_i f_i - J_5 u_{y} + J_6 u_{t} + J_8, ~~~ F_2= J_i g_i + J_5 u_{x} - J_7 u_{t} + J_9, ~~~
F_3=J_i h_i - J_6 u_{x} + J_7 u_{y} + J_{10}.
$$
Although the constants $J_5 - J_{10}$ give trivial contribution  to  conservation laws, they do effect non-trivially the characteristic condition, $Fg^{-1}F^t=0$,
where $g=f_{ij}$ is the $3\times 3$ symmetric matrix of the corresponding principal symbol. The characteristic condition imposes a system of quadratic constraints for the coefficients $J_1, \dots, J_{10}$ which,  in linearly degenerate integrable  cases,   specify a Veronese threefold $V^3\subset P^9$. For 3D equations of the form (\ref{u}), the concept of linear degeneracy can be defined as follows.
Looking for travelling wave solutions in the form
$ u(x, y, t)=u(\xi, \eta)+\zeta$ where $\xi, \eta, \zeta$ are arbitrary linear forms in the variables $x,y,t$, we obtain a  second-order PDE for $u(\xi, \eta)$,
$$
au_{\xi\xi}+2bu_{\xi\eta}+cu_{\eta \eta}=0,
$$
where the coefficients $a, b, c$ are certain functions of $u_{\xi}$ and $u_{\eta}$. Setting $v=u_{\xi}, \ w=u_{\eta}$, one can rewrite this PDE as a two-component system of hydrodynamic type. We say that Equation (\ref{u}) is linearly degenerate if all its travelling wave reductions are linearly degenerate in the sense of Sect. 2. The condition of linear degeneracy is equivalent to the identity (set $u_x,u_y,u_t=p_1,p_2,p_3$ and consider $f_{ij}$ as functions of $p_1, p_2, p_3$):
\begin{equation}
f_{(ij, k)}=c_{(k}f_{ij)},
\label{lind}
\end{equation}
here $f_{ij,k}=\partial_{p_k}f_{ij}$, $c_k$ is a covector, and brackets denote complete symmetrisation in $i, j, k$. 
Linearly degenerate integrable PDEs of the form (\ref{u}) 
were classified in \cite{FerMoss}:

\begin{theorem}    The following examples constitute a complete list  of linearly degenerate integrable PDEs:
$$
\mu u_{t}u_{xy}+\nu u_{y} u_{xt}+\eta u_{x}u_{yt}=0, ~~~ \mu+\nu+\eta=0,
$$
$$
u_{xx}+u_{x}u_{yt}-u_{y}u_{xt}=0,
$$
$$
u_{xy}+u_{y}u_{xt}-u_{x}u_{yt}=0,
$$
$$
u_{yy}+u_{xt}+u_{y} u_{tt}-u_{t} u_{yt}=0,
$$
$$
u_{xt}+u_{x}u_{yy}-u_{y}u_{xy}=0,
$$
$$
u_{tt}-u_{xx}-u_{yy}=0.
$$

\end{theorem}

In different contexts, the  canonical  forms of Theorem 1  have appeared  in \cite{Zakharevich, Pavlov, Shabat, Adler, Dun, Odesskii,  Manakov1, Manakov3, Ovsienko, Morozov}.  

The main result of this section is the following.

\begin{theorem}
\

\noindent (i) If a 3D quasilinear PDE of the form (\ref{u}) possesses `sufficiently many' characteristic integrals, then it must be linearly degenerate. Here `sufficiently many' means that the corresponding vector $F$  satisfies no extra algebraic constraints other than the characteristic condition itself, $Fg^{-1}F^t=0$.

\noindent (ii) Any  linearly degenerate integrable PDE (\ref{u}) possesses $V^3$-worth of characteristic integrals.

\end{theorem}

\medskip

\centerline{\bf Proof:}

\medskip

\noindent To demonstrate (i) we recall the result of \cite{B} according to which the functions $F_i$ defining a conservation law must satisfy the identity $F_{(i, j)}=sf_{ij}$, where $F_{i, j}=\partial_{p_j}F_i$, brackets denote symmetrisation in $i, j$, and $s$ is a coefficient of proportionality (all entries are viewed as functions of $p$'s). The characteristic constraint takes the form
$$
(f^{-1})^{ij}F_iF_j=0,
$$
 which can be rewritten as $f_{ij}F^iF^j=0$ where we use the notation $F_i=f_{ij}F^j$. Differentiating the characteristic condition by $p_k$ we obtain
$$
-(f^{-1})^{ip}f_{pq, k}(f^{-1})^{qj}F_iF_j+2(f^{-1})^{ij}F_{i, k}F_j=0,
$$
which can be rewritten as
$$
f_{pq, k}F^pF^q=2F_{i, k}F^i.
$$
Contracting this identity with $F^k$, using the condition $F_{(i, j)}=sf_{ij}$ and the characteristic constraint $f_{ij}F^iF^j=0$ we obtain the additional algebraic condition
\begin{equation}
f_{ij, k}F^iF^jF^k=0.
\label{cub}
\end{equation}
The requirement that this condition is satisfied identically modulo the characteristic constraint, $f_{ij}F^iF^j=0$, is equivalent to saying that the cubic (\ref{cub}) is divisible by the quadric $f_{ij}F^iF^j=0$,
$$
f_{ij, k}F^iF^jF^k=(c_iF^i)(f_{ij}F^iF^j),
$$
for some linear form $c_iF^i$. Symmetrisation of this identity implies the condition of linear degeneracy (\ref{lind}).

\medskip

Finally, the proof of (ii) is a case-by-case calculation. Details  are included in the Appendix. This finishes the proof of Theorem 2.

\medskip

Note that linearly non-degenerate or non-integrable equations may also possess characteristic integrals (alhough not `as many' as linearly degenerate integrable ones).

\medskip

\noindent {\bf Example 3}. The following integrable (linearly non-degenerate) equation,
$$
u_tu_{xy}+u_yu_{xt}+u_xu_{yt}=0,
$$
admits only three characteristic conservation laws:
$$
(u_x^2u_t)_y+(u_x^2u_y)_t=0,
$$
$$
(u_y^2u_t)_x+(u_y^2u_x)_t=0,
$$
$$
(u_t^2u_y)_x+(u_t^2u_x)_y=0.
$$

\section{Concluding remarks}

We have demonstrated that linearly degenerate second-order quasilinear integrable PDEs in 3D possess characteristic integrals parametrised by points on the Veronese variety $V^3$. Our calculations suggest that similar phenomena take place for other classes of linearly degenerate dispersionless integrable systems: all of them possess nontrivial characteristic integrals, what may change is the dimension of the corresponding Veronese variety.

\section{Appendix: characteristic integrals for equations from Theorem 1}

Here we present characteristic conservation laws for all examples from Theorem 1. For each of the  canonical forms, we present  four nontrivial conservation laws, and parametric expressions for the corresponding constants
$J_1, \dots, J_{10}$ as defined in Sect. 3. In each case these parametric equations are readily seen to specify a Veronese threefold.

\medskip

\noindent {\bf Equation 1} (discussed in the Introduction):
$$
\mu u_tu_{xy}+\nu u_y u_{xt}+\eta u_xu_{yt}=0.
$$
Four conservation laws:
$$
\eta (u_yu_t)_x+\nu(u_xu_t)_y+\mu(u_xu_y)_t=0,
$$
$$
\nu \left(\frac{u_y}{u_t}\right)_x+\eta \left(\frac{u_x}{u_t}\right)_y=0,
$$
$$
\mu \left(\frac{u_t}{u_y}\right)_x+\eta \left(\frac{u_x}{u_y}\right)_t=0,
$$
$$
\mu \left(\frac{u_y}{u_x}\right)_y+\nu \left(\frac{u_y}{u_x}\right)_t=0.
$$
Coefficients of characteristic integrals:
$$
J_1=\alpha^2, \hspace{3mm} J_2=\frac{1}{4 \nu \eta}\beta^2, \hspace{3mm} J_3=\frac{1}{4 \eta \mu}\delta^2, \hspace{3mm} J_4=\frac{1}{4 \nu \mu}\gamma^2, 
$$
$$
 J_5=\alpha \beta, \hspace{3mm} J_6=\alpha \delta, \hspace{3mm} J_7=\alpha \gamma, \hspace{3mm} J_8=-\frac{1}{2 \eta} \beta \delta, \hspace{3mm} J_9=-\frac{1}{2 \nu}\beta \gamma, \hspace{3mm} J_{10}=-\frac{1}{2\mu}\delta\gamma.
$$

\medskip

\noindent {\bf Equation 2.} 
$$
u_{xx}+u_xu_{yt}-u_yu_{xt}=0.
$$
Four conservation laws:
$$
\left(\frac{u_y}{2u_x^2}\right)_x+\left(\frac{1}{2u_x}\right)_y-\left(\frac{u_y^2}{2u_x^2}\right)_t=0,
$$
$$
(u_x-u_yu_t)_x+(u_xu_t)_y=0,
$$
$$
(2u_xu_t-u_yu_t^2)_x+(u_t^2u_x)_y-(u_x^2)_t=0,
$$
$$
-\left(\frac{1}{u_x}\right)_x+\left(\frac{u_y}{u_x}\right)_t=0.
$$
Coefficients of characteristic integrals:
$$
J_1=\alpha^2, \hspace{3mm} J_2=-\delta \beta, \hspace{3mm} J_3=\frac{1}{2} \beta^2, \hspace{3mm} J_4=\alpha \gamma, 
$$
$$
J_5=\frac{1}{2} \delta^2, \hspace{3mm}  J_6=\beta \gamma, \hspace{3mm} J_7=\alpha \beta, \hspace{3mm} J_8=-\alpha \beta - \gamma \delta, \hspace{3mm} J_9=\alpha \delta, \hspace{3mm} J_{10}=-\frac{1}{2}\gamma^2.
$$

\medskip

\noindent {\bf Equation 3.} 
$$
u_{xy}+u_yu_{xt}-u_xu_{yt}=0.
$$
Four conservation laws:
$$
(u_yu_t)_x+(u_x-u_xu_t)_y=0,
$$
$$
(u_yu_t^2)_x+(2u_xu_t-u_xu_t^2-u_x)_y-(u_xu_y)_t=0,
$$
$$
\left(\frac{1}{u_y}\right)_x-\left(\frac{u_x}{u_y}\right)_t=0,
$$
$$
-\left(\frac{1}{u_x}\right)_y-\left(\frac{u_y}{u_x}\right)_t=0.
$$
Coefficients of characteristic integrals:
$$
J_1=\alpha \beta, \hspace{3mm} J_2=\alpha^2, \hspace{3mm} J_3=\frac{1}{4} \delta^2, \hspace{3mm} J_4=\frac{1}{4}\gamma^2, 
$$
$$
J_5=-\frac{1}{4} \beta^2, \hspace{3mm}  J_6=-\alpha \delta, \hspace{3mm} J_7=\alpha \gamma, \hspace{3mm} J_8=-\frac{1}{2} \beta \delta, \hspace{3mm} J_9=\alpha \gamma - \frac{1}{2} \beta \gamma, \hspace{3mm} J_{10}=-\frac{1}{2}\delta\gamma.
$$

\medskip

\noindent {\bf Equation 4.} 
$$
u_{yy}+u_{xt}+u_yu_{tt}-u_tu_{yt}=0.
$$
Four conservation laws:
$$
(u_y-u_t^2)_y+(u_x+u_yu_t)_t=0,
$$
$$
(\frac{1}{2}u_t^2)_x+(u_yu_t-\frac{1}{2}u_t^3)_y+(-u_yu_t^2-\frac{1}{2}u_y^2+\frac{3}{2}u_t^2u_y)_t=0,
$$
$$
(u_yu_t-u_t^3)_x+(-u_xu_t+u_y^2-3u_yu_t^2+u_t^4)_y+(u_xu_y+2u_tu_y^2-u_yu_t^3)_t=0,
$$
$$
(u_y^2-2u_yu_t^2+u_t^4)_x+(-2u_xu_y+2u_xu_t^2-3u_y^2u_t+4u_yu_t^3-u_t^5)_y+(-2u_xu_yu_t-u_x^2-3u_y^2u_t^2+u_yu_t^4+u_y^3)_t=0.
$$
Coefficients of characteristic integrals:
$$
J_1=-\alpha \delta - \frac{1}{2} \beta \gamma, \hspace{3mm} J_2=2\alpha \gamma + \frac{1}{2}\delta^2, \hspace{3mm} J_3=\frac{1}{2}\delta\gamma, \hspace{3mm} J_4=\frac{1}{4}\gamma^2, 
$$
$$
J_5=\alpha\gamma,\hspace{3mm} J_6=-\alpha\delta, \hspace{3mm} J_7=\alpha^2+\frac{1}{2}\beta\delta, \hspace{3mm} J_8=\alpha^2 \hspace{3mm} J_9=\alpha\beta, \hspace{3mm} J_{10}=-\frac{1}{4}\beta^2.
$$

\medskip

\noindent {\bf Equation 5.} 
$$
u_{xt}+u_xu_{yy}-u_yu_{xy}=0.
$$
Four conservation laws:
$$
\left(\frac{u_y}{u_x}\right)_y-\left(\frac{1}{u_x}\right)_t=0,
$$
$$
(-u_y^2)_1+(u_xu_y)_y+(u_x)_t=0,
$$
$$
(-u_t^2+2u_tu_y^2-u_y^4)_x+(-2u_xu_yu_t+u_xu_y^3)_y+(u_xu_y^2)_t=0,
$$
$$
(u_tu_y-u_y^3)_x+(-u_xu_t+u_xu_y^2)_y+(u_xu_y)_t=0.
$$
Coefficients of characteristic integrals:
$$
J_1=\alpha^2, \hspace{3mm} J_2=-\frac{1}{2}\beta\gamma-\frac{1}{4}\delta^2, \hspace{3mm} J_3=-\frac{1}{4}\beta^2, \hspace{3mm} J_4=-\frac{1}{2}\beta\delta, 
$$
$$
J_5=-\frac{1}{2}\gamma\delta, \hspace{3mm} J_6=-\frac{1}{2}\beta\gamma, \hspace{3mm} J_7=\beta\alpha, \hspace{3mm} J_8=\frac{1}{4}\gamma^2, \hspace{3mm} J_9=\gamma\alpha, \hspace{3mm} J_{10}=\delta\alpha.
$$


\medskip






\begin{thebibliography}{99}









\bibitem{Adler} V.E. Adler and  A.B. Shabat, 
Model equation of the theory of solitons,  Theoret. and Math. Phys.  {\bf 153}, no. 1 (2007) 1373--1387.



\bibitem{Athorne95} C. Athorne,  A $Z^2\times R^3$ Toda system, Phys. Lett. A {\bf 206}, no. 3-4 (1995) 162--166. 


\bibitem{Athorne12} C. Athorne and H.  Yilmaz,  Laplace invariants for general hyperbolic systems,  J. Nonlinear Math. Phys. {\bf 19}, no. 3 (2012) 1250024, 20 pp.


\bibitem{B}  P.A. Burovskii, E.V. Ferapontov and S.P. Tsarev, Second order quasilinear PDEs and conformal structures in projective space, International J. Math. { \bf 21}, no. 6 (2010) 799--841.




\bibitem{Dun} M. Dunajski, A class of Einstein-Weyl spaces associated to an integrable system of hydrodynamic type. J. Geom. Phys. {\bf 51}, no. 1 (2004) 126--137.

\bibitem{Dun1} M. Dunajski and W. Krynski,
Einstein-Weyl geometry, dispersionless Hirota equation and Veronese webs, arXiv:1301.0621.







\bibitem{lagrange} E.V. Ferapontov, K.R. Khusnutdinova and S.P. Tsarev, On a class of three-dimensional integrable Lagrangians,  Comm. Math. Phys. {\bf 261}, no. 1 (2006)  225--243.



\bibitem{FerMoss} E.V. Ferapontov and J. Moss, Linearly degenerate PDEs and quadratic line complexes, arXiv:1204.2777.  






\bibitem{Hartley} D. Hartley, Darboux integrability in more than two dimensions, AAECC {\bf 11} (2001) 397--416.

\bibitem{Kakie75} K. Kaki\'e, 
On involutive systems of partial differential equations whose characters of order more than one vanish, 
Proc. Japan Acad. {\bf 51}, no. 4 (1975) 265--269. 

\bibitem{Liu} T.P. Liu, Development of singularities in the nonlinear waves for quasi-linear hyperbolic PDEs, J. Diff. Eq. {\bf 33} (1979) 92--111.	


\bibitem{Majda} A. Majda, Compressible fluid flows and systems of conservation laws in several space variables, Appl. Math. Sci., Springer-Verlag, NY, {\bf 53} (1984) 159 pp.

\bibitem{Manakov1} S.V. Manakov and P.M.  Santini,  Inverse scattering problem for vector fields and the Cauchy problem for the heavenly equation, Phys. Lett. A {\bf 359}, no. 6 (2006) 613--619.




\bibitem{Manakov3} S.V. Manakov and P.M.  Santini, 
On the solutions of the second heavenly and Pavlov equations, J. Phys. A {\bf 42}, no. 40 (2009) 404013, 11 pp.


\bibitem{Shabat} L. Martinez Alonso and A.B. Shabat, Hydrodynamic reductions and
solutions of a universal hierarchy, Theoret. and Mat. Phys. {\bf 140}, no. 2 (2004) 1073--1085.





\bibitem{Morozov} O. I. Morozov,
Recursion Operators and Nonlocal Symmetries for Integrable rmdKP and rdDym Equations,  arXiv:1202.2308.



\bibitem{Odesskii} A. Odesskii and V. Sokolov, Integrable (2+1)-dimensional systems of hydrodynamic type, 
Theoretical and Mathematical Physics {\bf 163}, no. 2 (2010) 549--586.

\bibitem{Ovsienko} V. Ovsienko, 
Bi-Hamiltonian nature of the equation $u_{tx}=u_{xy} u_y-u_{yy} u_x$, Adv. Pure Appl. Math. {\bf 1}, no. 1 (2010) 7--17.



\bibitem{Pavlov} M.V. Pavlov,  Integrable hydrodynamic chains, J. Math. Phys. {\bf 44} (2003) 4134--4156.

\bibitem{Parsons}  D.H. Parsons, 
The extension of Darboux's method. M\'emor. Sci. Math., Fasc. 142. Gauthier-Villars, Paris (1960) 75 pp. 



\bibitem{R1} B.L. Rozdestvenskii and A.D. Sidorenko, On the impossibility of `gradient catastrophe' for weakly nonlinear systems, Z. Vycisl. Mat. i Mat. Fiz. {\bf 7} (1967) 1176-1179.

\bibitem{R2} B.L. Rozdestvenskii and N.N. Yanenko, Systems of quasilinear equations and their applications to gas dynamics, translated from the second Russian edition by J. R. Schulenberger, Translations of Mathematical Monographs, {\bf 55} American Mathematical Society, Providence, RI (1983) 676 pp. 

\bibitem{Serre} D. Serre, Systems of conservation laws. 1.  Hyperbolicity, entropies, shock waves, Cambridge University Press,  (1999) 263 pp;  Systems of conservation laws. 2.  Geometric structures, oscillations, and initial-boundary value problems,  Cambridge University Press (2000) 269 pp. 

\bibitem{Sokolov} V.V. Sokolov and A.V.  Zhiber, On the Darboux integrable hyperbolic equations, Phys. Lett. A {\bf 208}, no. 4-6 (1995) 303--308.


\bibitem{Tsarev1} S. P. Tsarev, Generalized Laplace transformations and integration of hyperbolic systems of
linear partial differential equations, in Proc. Int. Symp. Symbolic and Algebraic Computation,
ISSAC 2005 (Beijing, China, July 24--27, 2005) (ACM Press, 2008), pp. 325--331.


\bibitem{Zakharevich} I. Zakharevich,
Nonlinear wave equation, nonlinear Riemann problem, and the twistor transform of Veronese webs, arXiv:math-ph/0006001.

\bibitem{Zhiber} A. V. Zhiber and V. V. Sokolov, Exactly integrable hyperbolic equations of Liouville type, Uspekhi Mat. Nauk {\bf 56} (2001) 63--106.







\end{thebibliography}
\end{document}